\documentclass[aps,prd,onecolumn,groupedaddress,showpacs,nofootinbib,amssymb]{revtex4}
\usepackage[T1]{fontenc}
\usepackage[latin1]{inputenc}
\usepackage{graphicx}
\usepackage[english]{babel}
\usepackage{amsmath}
\usepackage{amssymb}
\usepackage{amsfonts}

\begin{document}

\def\pp{{\, \mid \hskip -1.5mm =}}
\def\cL{{\cal L}}
\def\be{\begin{equation}}
\def\ee{\end{equation}}
\def\bea{\begin{eqnarray}}
\def\eea{\end{eqnarray}}
\def\beq{\begin{eqnarray}}
\def\eeq{\end{eqnarray}}
\def\tr{{\rm tr}\, }
\def\nn{\nonumber \\}
\def\e{\mathrm{e}}

%\begin{center}
%\Large\bf ???
%\end{center}

%\begin{center}
%{\Large
%Alexey Toporensky\\}
%{\large Shternberg Astronomical Institute,
%Moscow State University, Moscow, Russia\\}
%lesha@sai.msu.ru,
%\end{center}

%\begin{center}
%{\Large
%Petr Tretyakov\\}
%{\large Joint Institute for Nuclear Research, Dubna, Russia\\}
%tpv@xray.sai.msu.ru, tpv@theor.jinr.ru
%\end{center}

\title{Reconstruction and deceleration-acceleration transitions in
modified gravity}

\author{Shin'ichi Nojiri$^1$,
Sergei D. Odintsov$^{2,3}$\footnote{Also at TSPU, Tomsk},
Alexey Toporensky$^4$, and Petr Tretyakov$^5$}

\affiliation{
$^1$Department of Physics, Nagoya University, Nagoya 464-8602,
Japan \\
$^2$Instituci\`{o} Catalana de Recerca i Estudis Avan\c{c}ats
(ICREA), Barcelona \\
$^3$ Institut de Ciencies de l'Espai (IEEC-CSIC),
Campus UAB, Facultat de Ciencies, Torre C5-Par-2a pl, E-08193
Bellaterra (Barcelona), Spain \\
$^4$ Shternberg Astronomical Institute, Moscow State University,
Moscow, Russia \\
$^5$ Joint Institute for Nuclear Research, Dubna, Russia
}

\begin{abstract}

We discuss the cosmological reconstruction in modified Gauss-Bonnet
and $F(R)$ gravities.
Two alternative representations of the action (with and without
auxiliary scalar) are considered.
The approximate description of deceleration-acceleration transition
cosmologies is reconstructed.
It is shown that cosmological solution containing Big Bang and Big
Rip singularities may be reconstructed only using the representation
with the auxiliary field.
The analytical description of the deceleration-acceleration
transition cosmology in modified Gauss-Bonnet gravity is demonstrated
to be impossible at sufficiently general conditions.

\end{abstract}

\pacs{95.36.+x, 98.80.Cq, 04.50.Kd, 11.10.Kk, 11.25.-w}

\maketitle

\section{Introduction}

The modified gravity approach (for general review, see \cite{review})
became the essential element of the modern cosmology.
It is quite remarkable that some change of the classical
gravitational action
may resolve the number of cosmological problems, including
inflationary paradigm, dark energy and dark matter. It turns out that
it is not necessary to introduce the extra ingredients (usually,
scalar or fluid) as all these phenomena could be understood as
gravitational manifestations.
For instance, the unification of the early-time inflation and
late-time acceleration may be achieved in $F(R)$ gravity (for first
realistic model of that sort, see \cite{NO}) without the need to
introduce the inflaton and (scalar) dark energy by hands.
Several models of modified gravity may successfully describe dark
matter as gravitational effect (for a recent review,
see \cite{faraoni}).
The coincidence problem effectively disappears in the modified
gravity approach because dark matter and dark energy are caused by
the
universe expansion governed by specific theory. It is expected that
modified gravity may be
helpful also in high-energy physics (for instance, for hierarchy
problem).

Unfortunately, the realistic modified gravity has usually highly
non-linear structure in terms of geometric invariants (curvature,
Gauss-Bonnet invariant, etc.). As the result, its background evolution
is very hard to describe analytically unlike to the case of General
Relativity where number of viable analytic solutions are available.
In turn, with only approximate FRW solutions of modified gravity it
is extremely difficult to study the cosmological perturbations. At
best, such cosmological perturbations are studied in further
approximation neglecting the higher-derivatives non-linearities which
is definitely not sufficient.
In order to study the background evolution of the alternative
gravities, so-called reconstruction method has been developed (for
the introduction, see \cite{Nojiri:2006be}). Within the
reconstruction method,
given FRW cosmology may be used to reconstruct the modified gravity
where such cosmology is the solution of the equations of motion.

In the present paper we develop the reconstruction method for
modified Gauss-Bonnet gravity \cite{FG}. It is demonstrated how to
reconstruct the
theory which admits the deceleration-acceleration transition (the
transition to $\Lambda$CDM epoch). Such background evolution turns
out to be the very complicated and approximate one.
For quite general class of $F(G)$-functions we show that there is no
analytical description of deceleration-acceleration transition. It
turns out that it is very difficult (if possible at all) to construct
such a model which admits such transition analytically (the
co-existence of matter dominance and accelerating
solutions \cite{Dunsby}).
The comparison with $F(R)=R+f(R)$ theory is done. The alternative
presentation for $F(R)$
and $F(G)$ modified gravity using the auxiliary scalar is considered.
It is shown that reconstruction using such representation leads to
wider class of cosmological solutions, including the
deceleration-acceleration transition ones.
It is demonstrated
that cosmological solution containing the Big Bang as well as Big Rip
singularity may be reconstructed from $F(R)$ gravity.

\section{Analytical approach to deceleration-acceleration transition
in $f(G)$-gravity}

Let us study modified gravity with the following action \cite{FG}:
\begin{equation}
S_{F(G)}=\int d^4 x \sqrt{-g} \left(
\frac{R}{2\kappa^2}+F(G)+\mathcal{L}_{\mathrm{matter}}\right)\, .
\label{1.1}
\end{equation}
Here $G$ is the Gauss-Bonnet invariant
$G=R^2-4R_{\mu\nu}R^{\mu\nu}+R_{\mu\nu\rho\sigma}R^{\mu\nu\rho\sigma}$
and $\mathcal{L}_{\mathrm{matter}}$ is the Lagrangian of matter. It
is convenient to put $2\kappa^2=1$ in this section.
We will discuss only the FRW background:
\begin{equation}
g_{\mu\nu}=diag(-n^2,a^2,a^2,a^2)\, .
\label{1.2}
\end{equation}
The variation of the action (\ref{1.1}) with respect to
lapse-function $n$ gives the modified Friedman equation:
\begin{equation}
6H^2+F-F'G+24F''H^3\dot G=\rho\, ,
\label{1.3}
\end{equation}
and the variation with respect to scale factor $a$ gives more
complicated equation:
\begin{equation}
4\dot H +6H^2 + F - F'G + 8 F''' H^2\dot G^2 +\frac{2G}{3H} F'' \dot
G +8H^2F'' \ddot G = -p\, .
\label{1.4}
\end{equation}
Here $\rho$ and $p$ is the matter energy-density and pressure,
respectively, which
arises from $\mathcal{L}_{\mathrm{matter}}$. We also note that prime
$'$ denotes partial differentiation
of function $F$ with respect to its argument.
Using analogy with the FRW equations in the Einstein gravity
one may define $\rho_G \equiv -F+F'G-24F''H^3\dot G$
and $p_G \equiv F - F'G + 8 F''' H^2\dot G^2
+\frac{2G}{3H} F'' \dot G +8H^2F'' \ddot G$,
so the equations (\ref{1.3}) and (\ref{1.4}) take the following form:
\begin{equation}
6H^2=\rho_\mathrm{tot}\, ,
\label{1.5}
\end{equation}
\begin{equation}
4\dot H +6H^2 = -p_\mathrm{tot}\, ,
\label{1.6}
\end{equation}
where $\rho_\mathrm{tot}=\rho+\rho_G$ and $p_\mathrm{tot}=p+p_{G}$.
Note that different cosmological solutions for above theory have been
discussed in refs.\cite{2}.

The barotropic equation of matter state $p=w\rho$ is considered
below.
To close our system one adds conservation energy law which takes the
following form:
\begin{equation}
\dot\rho + 3H(\rho+p)=0\, .
\label{1.7}
\end{equation}
Note also that equation (\ref{1.3}) is just the first integral of the
system (\ref{1.4})-(\ref{1.7}).
Now let us consider the possibility of occurrence of late-time
universe acceleration due to function $F(G)$ analytically.
In other words, one searches some function $F(G)$, which plays
non-essential role during the standard dust stage($w=0$),
but gives leading contribution at late times. Our purpose is the
analytical description of such deceleration-acceleration transition.
We discuss functions with $F(0)=0$, because otherwise we will have
some analogue of cosmological constant.
(Of course, permitting the effective cosmological constant may
qualitatively change the results obtained below).
The effective equation of state parameter may be easily found by
using the expressions (\ref{1.5})-(\ref{1.6}):
$w_\mathrm{eff}=-1-\frac{2\dot H}{3H^2}$.

Now let us suppose that there exists some function $F(G)$ which leads
to late-time acceleration and has the following properties:
\begin{equation}
F(0)=0,\, F'(0)=0,\, F''(0)=0,\, F'''(0)=0\, .
\label{1.8}
\end{equation}
 From another side it is known that at the deceleration-acceleration
transition point $G=24\frac{\dot a^2\ddot a}{a^3}=0$.
This point is reached when $w_\mathrm{tot}=-1/3$. By calculating the
values of $\rho_G$ and $p_G$ at the transition
point, we find that they vanish. This means that there is no any
effective matter besides the usual matter $\rho$
at this moment, so $\rho_\mathrm{tot}=\rho$ and $p_\mathrm{tot}=p$,
hence $w_\mathrm{tot}$ must be equal to some value of $w$ which is
bigger than $-1/3$. This logical contradiction proves that any
function satisfying the conditions (\ref{1.8})
cannot reach the deceleration-acceleration transition point.
Note also that the condition $F'(0)=0$ may be removed from
(\ref{1.8}) because its contribution to $\rho_G$ and $p_G$
contains $G$ as a factor.
This result is complimentary to the one of ref.\cite{Dunsby}
where it has been shown that some class of $F(G)$-theories which
allow an exact power-law
solution can not explain transition from deceleration to
acceleration.
Actually, the following general form
of function which allows \textit{exact} power-law decelerating
solution (see (15) in \cite{Dunsby}) is: $F(G)=AG^{0.5}+BG^k$
where $k<\frac{3}{2}$ and this function does not satisfy to our
condition (\ref{1.8}). From another side, one may easily
find functions which satisfy the conditions (\ref{1.8}) but do not
allow \textit{exact} power-law solution.
For example,
\begin{equation}
F=\sum_{N=4}^{\infty} a_NG^N\, ,\quad F=\frac{G^N}{c_1G^N+c_2}\, .
\label{1.9}
\end{equation}
The situation is the following: there is some function $F$,
which does not allow exact power-law decelerating solution, but
allows it approximately with very good accuracy.
This solution may be unstable and leading to acceleration. The
examples of such approximate deceleration-acceleration transition
will be discussed below.

\section{Comparison with $F(R)$-gravity}

It is interesting to compare results of the previous section with
$F(R)$-gravity.
Its action has the following form:
\begin{equation}
S_{f(R)}=\int d^4 x \sqrt{-g} \left (\frac{1}{2\kappa^2}
\left[R+f(R)\right]+\mathcal{L}_{\mathrm{matter}} \right)\, .
\label{3.1}
\end{equation}
FRW equations of motion are (here again $2\kappa^2=1$):
\begin{equation}
6H^2+f-f'R + 6H^2f'+6Hf''\dot R=\rho\, ,
\label{3.2}
\end{equation}
and
\begin{equation}
4\dot H +6H^2 + f - f'R + 6H^2f' + 2 f'''\dot R^2 + 4H f'' \dot R
+ 2 f'' \ddot R = -p\, .
\label{3.3}
\end{equation}
The latter equation is a consequence of (\ref{3.2}) and (\ref{1.7}).
It is well known that $R=0$ identically for the regime
$a\sim t^{1/2}$,
which corresponds to $w_\mathrm{eff}=\frac{1}{3}$.
So using developed analysis one may try to investigate the
possibility to reach $a\sim t^{1/2}$ regime.
We will study only functions satisfying the conditions
\begin{equation}
f(0)=0,\, f'(0)=0,\, f''(0)=0,\, f'''(0)=0\, .
\label{3.4}
\end{equation}

Let us consider the theory with fixed EoS $w_m$ matter besides the
relativistic matter ($w_m=\frac{1}{3}$). Following previous section
we define $\rho_{f(R)}\equiv-f+f'R - 6H^2f'-6Hf''\dot R$ and
$p_{f(R)}\equiv f - f'R + 6H^2f' + 2 f'''\dot R^2 + 4H f'' \dot R + 2
f'' \ddot R$ to rewrite equations (\ref{3.2})-(\ref{3.3}) in the
canonical form (\ref{1.5})-(\ref{1.6}).
In this case we have the logical contradiction: from the one
side it must be $w_\mathrm{tot}=\frac{1}{3}$ on
$a\sim t^{1/2}$ regime, but from another side we have
$w_\mathrm{tot}=w_m\neq \frac{1}{3}$ because there
is no any contribution to $\rho_\mathrm{tot}$ and $p_\mathrm{tot}$
from $f(R)$-terms at this regime ($\rho_{f(R)}=0$, $p_{f(R)}=0$ due
to (\ref{3.4})).

So we have the following result. The analytical description of
transition to regime $a\sim t^{1/2}$
in the universe with any perfect fluid except $w_m=\frac{1}{3}$ in
$f(R)$-gravity with (\ref{3.4}) is very hard to realize
(compare with \cite{Dunsby1} where similar conclusion is made).
Of course, other classes of functions $f(R)$ or account of the
effective cosmological constant may change this conclusion.

Note also that there is no any problem with deceleration-acceleration
transition in $f(R)$ gravity. A number of such theories admitting the
transition is well known. For example, most general function which
leads from matter dominated era to the $\Lambda$CDM cosmology was
constructed in \cite{NOSG, 06} by using reconstruction method. Below
we try to use this method to solve the problem described in previous
section.

\section{Reconstruction and the deceleration-acceleration transition
in $F(G)$-gravity}

Let us investigate possibility to find the theories (\ref{1.1}) which
allow transition from deceleration
to acceleration phase by using reconstruction method. This method
developed in ref.\cite{NOSG} may be easily adopted
to our Gauss-Bonnet modified gravity (\ref{1.1}). We start from the
equation (\ref{1.3}).
First of all we would like to use a new variable $N$ instead of the
cosmological time $t$, defined by $N=\ln\frac{a}{a_0}$.
%%%%%%%%%%
Here $a_0$ is the value of the scale factor $a(t)$ in (\ref{1.2}) at a 
fixed time. 
%%%%%%%%%%
This variable is related with the redshift $z$ by
$\mathrm{e}^{-N}=1+z$. Since $\frac{d}{dt}=H\frac{d}{dN}$
and
$\frac{d^2}{dt^2}=H^2\frac{d^2}{dN^2}+H\frac{dH}{dN}\frac{d}{dN}$,
one can rewrite (\ref{1.3}) as
\begin{equation}
6H^2+F(G) - 24H^3(H'+H)F'(G) + 24^2 F''(G)H^6(HH''+3H'^2+4HH')=\rho
\, .
\label{2.1}
\end{equation}
Here $H'\equiv dH/dN$ and $H''\equiv d^2H/dN^2$, but $F'=dF/dG$ like
above. Here we have used
also $G=24H^2(\dot H+H^2)=24H^3(H'+H)$. If the matter energy density
$\rho$ is given by a sum of the fluid densities
with constant EoS parameter $w_i$, we find
\begin{equation}
\rho=\sum\rho_{i0}a^{-3(1+w_i)}=\sum\rho_{i0}a_0^{-3(1+w_i)}
\mathrm{e}^{-3(1 + w_i)N}\, .
\label{2.2}
\end{equation}
%%%%%%%%%%%%%%
Here $\rho_{i0}$ is a constant. 
%%%%%%%%%%%%
Let the Hubble rate is given in terms of $N$ via some function $k(N)$
as
\begin{equation}
H=k(N)=k(-\ln(1+z))\, .
\label{2.3}
\end{equation}
Note now that the expression $G = 24k(N)^3k'(N)+24k(N)^4$ may be
solved with respect to $N$ as $N=N(G)$.
Then by using (\ref{2.2}) and (\ref{2.3}), one can rewrite
(\ref{2.1}) as
\begin{eqnarray}
&& 6\left(k\left(N\left(G\right)\right)\right)^2 +F(G) 
 - 24\left(k\left(N\left(G\right)\right)\right)^3F'(G)\left[k'\left(N\left(G\right)\right)
+k\left(N\left(G\right)\right)\right] \nn
&& +24^2F''(G)\left(k\left(N\left(G\right)\right)\right)^6
\left[k\left(N\left(G\right)\right)k''\left(N\left(G\right)\right) 
+ 3\left(k'\left(N\left(G\right)\right)\right)^2
+ 4k\left(N\left(G\right)\right)k'\left(N\left(G\right)\right)\right]
\nn
&& = \sum\rho_{i0}a_0^{-3(1+w_i)}\mathrm{e}^{-3(1+w_i)N}\, .
\label{2.4}
\end{eqnarray}
This equation is differential equation for $F(G)$ and may be
simplified by introducing $h(N)\equiv \left(k\left(N\right)\right)^2=H^2$:
\begin{eqnarray}
&& 6h(N(G)) +F(G) - 12\frac{dF(G)}{dG}
\left[h(N(G))h'(N(G))+2\left(h\left(N\left(G\right)\right)\right)^2\right]
\nn
&& +24^2\frac{d^2F(G)}{dG^2}h\left(N\left(G\right)\right)^3
\left[\frac{1}{2}h''\left(N\left(G\right)\right) + 2 h'\left(N\left(G\right)\right)
+\frac{h'\left(N\left(G\right)\right)^2}{h \left(N\left(G\right)\right)}\right] \nn
&& = \sum\rho_{i0}a_0^{-3(1+w_i)}\mathrm{e}^{-3(1+w_i)N}\, .
\label{2.5}
\end{eqnarray}
Note that the Gauss-Bonnet invariant is given by
$G=24h(N)^2+12h(N)h'(N)$. Hence, when we find $F(G)$ satisfying
the differential equation (\ref{2.5}), such $F(G)$ theory admits the
solution (\ref{2.3}) and therefore such
gravity realizes above cosmological solution. This is essentially the
cosmological reconstruction.

Now let us discuss the simplest example which is related with
previous discussion and which reproduces the $\Lambda$CDM-era.
In the Einstein gravity the FRW equation for the $\Lambda$CDM
cosmology is given by
\begin{equation}
6H^2=6H_0^2+\rho_0a^{-3}=6H_0^2+\rho_0 a_0^{-3}\mathrm{e}^{-3N}\, .
\label{2.6}
\end{equation}
%%%%%%%%%%%%%%%
Here $H_0$ and $\rho_0$ are constants. 
%%%%%%%%%%%%%%%
This equation reproduces the universe with dust matter which enters
to $\Lambda$CDM-era
at late time (for sufficiently small $H_0$). Therefore, it reaches
the point $\ddot a=0$ at some moment. So we have
\begin{equation}
h(N)=H_0^2+\frac{1}{6}\rho_0 a_0^{-3}\mathrm{e}^{-3N}\, .
\label{2.7}
\end{equation}
Substituting this relation into expression for $G$ we find:
\begin{equation}
G(N)=24 H_0^4+2H_0^2\rho_0a_0^{-3}\mathrm{e}^{-3N}-\frac{1}{3}\rho_0^2a_0^{-6}
\mathrm{e}^{-6N}\, ,
\label{2.8}
\end{equation}
which may be solved to find $N(G)$. It is convenient to introduce
$x=\rho_0 a_0^{-3} \mathrm{e}^{-3N}$, so finally
we have
\begin{equation}
24 H_0^4 + 2H_0^2 x -\frac{1}{3}x^2 = G\, .
\label{2.9}
\end{equation}
It is interesting to note that $x>0$ at any moment. Moreover, one may
easily calculate the transition point
from deceleration to acceleration which corresponds to $G=0$:
$x(G=0)=12H_0^2$. The solution of (\ref{2.9}) is:
\begin{equation}
x_{1,2}=3H_0^2 \pm \sqrt{9^2H_0^4-3G}\, .
\label{2.10}
\end{equation}
Note that sign ``$-$'' must be excluded because it corresponds to
non-physical negative values of $x$
for negative $G$, which corresponds to accelerated regimes. Now we
can see that only values $G<27H_0^4$ are resolved.
Now it is necessary to use the function (\ref{2.7}) in order to find the
theory $F(G)$ as the solution of the differential equation (\ref{2.5}).
It turns out that this differential equation is extremely complicated
and the corresponding solution may be found only numerically for
different asymptotics (near to transition point). In principle, it is
easier to construct such solutions in the alternative representation
of $F(G)$ theory with auxiliary scalar. The corresponding examples
are found in third and fourth papers from ref.\cite{2}. That is why
we will no go further to technical details of the solution of
eq.(\ref{2.5}).
Hence, in principle it is possible to construct $F(G)$ which allows
the transition from deceleration to acceleration era
(of course if we can solve the corresponding differential equation).

\section{Alternative representations of $F(R)$-gravity and
$F(G)$-gravity and the reconstruction}

Let us discuss the alternative representation for $F(R)$-gravity and
$F(G)$-gravity with the actions given by (\ref{3.1}) and
(\ref{1.1}), respectively.

In addition to the problems mentioned in the previous sections, there
appear other problems in $F(R)$- and $F(G)$-gravities.
First problem is easy to understand in terms of $F(R)$ gravity.
The Einstein gravity coupled with perfect fluid with constant
equation of state (EoS) parameter $w$
can be reproduced by the following $F(R)$ theory
\be
\label{AI}
F(R) \propto R^m\, , \quad
m = \frac{9w + 7 \pm \sqrt{45 w^2 + 126 w + 53}}{6\left(w+1\right)}\
\mbox{or} \
w = - 1 - \frac{2(m-2)}{3(m-1)(2m-1)}\, .
\ee
We may investigate the modified gravity with Big Bang singularity
with $w_\mathrm{BB}>0$ and Big Rip singularity \cite{sing}
with $w=w_\mathrm{BR} <0$. In both of the Big Bang singularity and
Big Rip singularity, the
scalar curvature $R$ diverges. This shows that if we construct a
model describing both of
Big Bang singularity and Big Rip singularity, the corresponding
$F(R)$ must be double valued function of $R$.

Similarly for $F(G)$-gravity, if we try to construct a realistic
model, where there is a transition
from decelerating phase to the accelerating phase, $F(G)$ may become
a double valued function or it may become purely imaginary function.

In the following, we consider how the above problem could be solved.
At least locally we can rewrite the actions (\ref{3.1}) and
(\ref{1.1})
by introducing the auxiliary scalar field $\phi$ as follows
\be
\label{III}
\tilde S_{F(R)} = \int dx^4 \sqrt{-g} \left( \frac{P(\phi) + Q(\phi)
R}{2\kappa^2}
+ \mathcal{L}_\mathrm{matter} \right)\, ,
\ee
and
\be
\label{IV}
\tilde S_{F(G)} = \int dx^4 \sqrt{-g} \left( \frac{R}{2\kappa^2} -
V(\phi) + f(\phi)G
+ \mathcal{L}_\mathrm{matter} \right)\, .
\ee
We should note, however, the actions (\ref{III}) and (\ref{IV})
express more wide class of theories
than the actions (\ref{3.1}) and (\ref{1.1}) (for related discussion,
see also \cite{pons}).
For example, we may consider the following model corresponding to
$F(R)$ gravity:
\be
\label{V}
P(\phi) = \frac{1}{3} \phi^3 + \beta \phi^2 \, , \quad Q(\phi)
= \gamma \phi \, .
\ee
Here $\beta$ 
%%%%%%%%%%
and $\gamma$ are constants. 
%is a constant.
(The following arguments do apply even for $F(G)$ gravity.)
Then by the variation of $\phi$, one finds
\be
\label{VI}
0 = \phi^2 + 2 \beta \phi + \gamma R \, ,
\ee
which can be solved with respect $\phi$ as
\be
\label{VII}
\phi = - \beta \pm \sqrt{ \beta^2 - \gamma R}\ ,
\ee
which gives
\be
\label{VIII}
\tilde S_{F(R)} = \int dx^4 \sqrt{-g} \left( \frac{F_\pm
(R)}{2\kappa^2}
+ \mathcal{L}_\mathrm{matter} \right)\, ,\quad
F_\pm (R) \equiv \left( - \frac{2\beta^2}{3}
+ \frac{\gamma R}{3} \right) \left( - \beta \pm \sqrt{ \beta^2 -
\gamma R} \right)
\ee
The action (\ref{VIII}) is double-valued function and furthermore the
value of $R$ is restricted
to be $\gamma R< \beta^2$ in order to have the real
$\tilde S_{F(R)}$. Hence, the action (\ref{3.1}) describes the theory
corresponding to one of the branches of
double-valued function and $R$ is restricted to be $\gamma R<
\beta^2$.
We should note, however, that we need not to start from the action
(\ref{3.1}) but from the action
(\ref{III}). The action (\ref{III}) may describe the scalar field
theory with potential
$-\frac{P(\phi)}{2\kappa^2}$ and the Brans-Dicke non-minimal coupling
$\frac{Q(\phi)}{2\kappa^2}$
but without the kinetic term. If we start with the action
(\ref{III}) instead of (\ref{3.1}) from the very beginning, even if
we consider the model (\ref{V}), we may
obtain the theory with transition between $F_+(R)$ and $F_-(R)$. Note
that, in the model
corresponding to (\ref{III}) with (\ref{V}), the value of $R$ can be,
in general,
in the region $\gamma R< \beta^2$, which is forbidden for the action
(\ref{3.1}).

Let us clarify it in more detail. For simplicity, we neglect the
contribution from matter by omitting $\mathcal{L}_\mathrm{matter}$.
First we consider the following model corresponding to (\ref{III}):
\bea
\label{IX}
&& P(\phi) = \e^{\tilde{g}(\phi)/2} \tilde{p}(\phi)\ ,
\quad \tilde{g}(\phi) = - 10 \ln \left[
\left(\frac{\phi}{t_0}\right)^{-\gamma}
 - C \left(\frac{\phi}{t_0}\right)^{\gamma+1} \right]\ , \nn
&& \tilde{p}(\phi) = \tilde{p}_+ \phi^{\beta_+} + \tilde{p}_-
\phi^{\beta_-}\, ,
\quad \beta_\pm \equiv \frac{1 \pm \sqrt{1 + 100 \gamma (\gamma 
+ 1)}}{2}\, , \nn
&& Q(\phi) = -6 \left[ \frac{d \tilde{g}(\phi)}{d\phi} \right]^2
P(\phi)
-6\frac{d \tilde{g}(\phi)}{d\phi} \frac{dP(\phi)}{d\phi} \, ,
\eea
%%%%%%%%%%%
Here $t_0$, $C$, and $\tilde{p}_\pm$ are constants. 
%%%%%%%%%%%%
Now $P(\phi)$ and $Q(\phi)$ are smooth functions of $\phi$ as long as
\be
\label{IXb}
0<\phi< t_s \equiv t_0 C^{-1/(2\gamma + 1)}\, .
\ee
The exact solution of the FRW equation is
\be
\label{X}
H(t) = \left(\frac{10}{t_0}\right) \left[ \frac{ \gamma
\left(\frac{t}{t_0}\right)^{-\gamma-1 }
 + (\gamma+1) C \left(\frac{t}{t_0}\right)^{\gamma}
}{\left(\frac{t}{t_0}\right)^{-\gamma}
 - C \left(\frac{t}{t_0}\right)^{\gamma+1}}\right]\ ,
\ee
When $t\to 0$, i.e., $t \ll t_s$, $H(t)$ behaves as
\be
\label{PDF10}
H(t) \sim \frac{10\gamma}{t}\, ,
\ee
which corresponds to the Big Bang singularity at $t=0$.
On the other hand, when $t\to t_s$, we find
\be
\label{PDF12}
H(t) \sim \frac{10}{t_s - t}\, .
\ee
which corresponds to the Big Rip singularity.
Then in the form (\ref{III}) of the action, one can obtain the
cosmological model describing both of the Big Bang
and Big Rip singularities. In this alternative presentation with
auxiliary scalar, it is easy also to construct the
deceleration-acceleration transition solutions. Such reconstruction
has been presented already for $F(R)$ and $F(G)$ theories in
refs.\cite{06,Nojiri:2006be} and third and fourth papers from
ref.\cite{2}. That is why we do not give the details of such
cosmologically-viable theories here.

We should also note that the modified gravity which exhibits the
transition from deceleration epoch to acceleration
epoch can be obtained without introducing the auxiliary field $\phi$
(compare with reconstruction in refs.\cite{NOSG,06,dobado}). For
example, we consider the following form of $F(R)$:
\be
\label{RZ22}
F(x) = A F(\alpha,\beta,\gamma;x) + B x^{1-\gamma} F(\alpha - \gamma
+ 1, \beta - \gamma + 1, 2-\gamma;x)\ .
\ee
Here $A$ and $B$ are constants, $F(\alpha,\beta,\gamma;x)$ is Gauss'
hypergeometric function,
$x$ is defined by $x=\frac{R}{3H_0^2} - 3$, and
\be
\label{RZ20}
\gamma = - \frac{1}{2}\ ,\alpha + \beta = - \frac{1}{6}\ ,\quad
\alpha\beta = - \frac{1}{6}\ .
\ee
The action has an exact solution which reproduces, without real
matter, the $\Lambda$CDM era whose
FRW equation is given by
\be
\label{RZ13}
\frac{3}{\kappa^2} H^2 = \frac{3}{\kappa^2} H_0^2 + \rho_0 a^{-3}\, .
\ee

Next we consider the following $F(G)$ gravity model corresponding to
the action (\ref{IV}):
\bea
\label{XI}
V(\phi) &=& \frac{3}{\phi_0 \kappa^2}\left( 1 +
g_1\frac{\phi_0}{\phi} \right)^2
 - \frac{6g_1}{\phi_0^2 \kappa^2}\left( 1 + g_1\frac{\phi_0}{\phi}
\right)
\left(\frac{\phi}{\phi_0}\right)^{g_1} W\left( - g_1 - 1,
\frac{\phi}{\phi_0} \right)\, ,\nn
f(\phi) &=& \frac{\phi_0^2 g_1}{4\kappa^2}
\int^{\frac{\phi}{\phi_0}} dx
\frac{\e^x x^{g_1}}{\left( 1 + g_1 x \right)^2}
W\left( - g_1 - 1, x \right)\, .
\eea
Here $g_1$ and $\phi_0$ and positive constants and $\$W(\alpha,x)$ is
given by the incomplete gamma function:
\be
\label{XII}
W(\alpha,x) = \int^x dy \e^{-y} y^{\alpha - 1}\, .
\ee
Note that the functions $V(\phi)$ and $f(\phi)$ are smooth functions
as long as $\phi>0$.
An exact solution of the model is given by
\be
\label{XIII}
H(t) = \frac{1}{\phi_0} + \frac{g_1}{t}\, .
\ee
When $t$ is small $H(t)$ describes the Big Bang singularity where the
expansion of the universe
is decelerating if $g_1<1$. On the other hand, when $t$ is large $H$
goes to a constant:
$H\to \frac{1}{\phi_0}$, which corresponds to the de Sitter universe
and the universe is expanding with the acceleration..
Hence, starting from the theory with the action (\ref{IV}), one can
explicitly
construct a model which admits the approximate transition from
decelerating phase to the accelerating phase,

\section{Discussion}

In summary, we discussed the cosmological reconstruction method for
modified Gauss-Bonnet and $F(R)$ gravities.
Two alternative representations for the action is used: with and
without the auxiliary scalar field. It turns out that the
cosmological solutions in the representation with the auxiliary
scalar follow from the wider class of theories.
Moreover, it is easier to reconstruct modified gravity in such
representation.
For instance, the cosmological solution which contains the Big Bang
and Big Rip singularities may be reconstructed in such formulation
with the auxiliary scalar but not in the original formulation.
Special attention is paid to the cosmologies admitting the
deceleration-acceleration transitions. It is shown that such
cosmological solutions may be reconstructed in both representations of
modified gravity but only approximately. The analytical
deceleration-acceleration transition cosmology in modified
Gauss-Bonnet gravity satisfying to some reasonable conditions is
shown to be impossible. It is extremely hard (if possible at all) to
find such analytical solutions in modified Gauss-Bonnet gravity.

The detailed understanding of the background evolution of modified
gravity is the necessary step in the development of the cosmological
perturbations. Hence, even the approximate background evolution
realized via the reconstruction method may serve for this purpose in
order to select the most realistic theories confronting them with the
observational data.

\section*{Acknowledgments}

This work was partially supported by RFBR grant 08-02-00923 and
by the scientific school grant 4899.2008.2 of the Russian Ministry of
Science and Technology (PT and AT) and by RFBR grant 09-02-12417 (PT).
The work by S.N. is supported in part by Global
COE Program of Nagoya University provided by the Japan Society
for the Promotion of Science (G07).
The work by S.D.O. is supported in part by MICINN (Spain) project
FIS2006-02842 and by AGAUR (Generalitat de Catalunya), project 2009
SGR994.

\end{document}